\documentclass[preprint]{aastex}		

\usepackage{subfigure}
\usepackage{appendix}
\usepackage{booktabs}

\begin{document}
\title{Decontaminating {\it Swift} UVOT Grism Observations of Transient Sources}
\author{Michael~T.~Smitka$^{1}$, Peter~J.~Brown$^{1}$, Paul~Kuin$^{2}$, Nicholas~B.~Suntzeff$^{1}$}
\affil{$^{1}$George P. and Cynthia Woods Mitchell Institute for Fundamental Physics \& Astronomy, \\
Texas A\&M University, Department of Physics and Astronomy, \\
4242 TAMU, College Station, TX 77843, USA \\
$^{2}$ Mullard Space Science Laboratory/University College London \\
Holmbury St. Mary, Dorking, Surrey, RH5 6NT, UK}

\email{mikesmitka@gmail.com} 
\email{grbpeter@yahoo.com}
\email{n.kuin@ucl.ac.uk}
\email{nsuntzeff@tamu.edu}

\begin{abstract}

We present a new technique of decontaminating {\it Swift} UVOT grism spectra for transient objects.  We describe the template image requirements and image processing steps necessary to successfully implement the empirical decontamination technique.  We demonstrate the accuracy of the flux and wavelength calibrations for decontaminated spectra by comparing a spectrum of SN~2011fe with a well-calibrated, long-slit ultraviolet spectrum from the Hubble Space Telescope's Space Telescope Imaging Spectrograph.  We also show how the decontamination removes spurious emission lines from spectra of iPTF14bdn which otherwise could be misinterpreted as coming from the supernova.  The software which implements this technique is briefly discussed and is made available to the community.
 
\end{abstract}


\section{Introduction  \label{intro}}
The {\it Swift} Ultraviolet/Optical Telescope (UVOT) \citep{Gehrels_etal_2004, Roming_etal_2005} is well-suited for observing transient events in the ultraviolet (UV) because of its fast targeting response and UV sensitivity.  For spectroscopy the UVOT instrument features two slitless grism options: the UV-grism ($1700-5000 $\,\AA) and the V-grism ($2850-6600$\,\AA) \citep{Kuin2015}.  In practice, the UV-grism is heavily favored because it is sensitive to a larger range of wavelengths not accessible to ground-based observatories.  {\it Swift's} fast targeting and flexible scheduling are inherent to its primary mission as a gamma-ray burst observatory and give the UVOT instrument the ability to target objects with just a few hours notice.  These qualities make the UVOT instrument ideal for studying the early evolution of transient objects in the UV, which are otherwise difficult or impossible to observe.  

The slitless design of the UVOT grisms makes the contamination of spectra a major concern.  In a slitless grism design, a particular feature in a spectrum depends both on the wavelength and the position of the object on the detector.  When the spectra of a transient object are contaminated, the data are oftentimes completely lost because the target cannot be reobserved.  In our analysis of Type Ia supernovae (SNe Ia) UVOT UV-grism spectra, we determined that roughly half of the available observations showed some signs of contamination and were thus partially or completely compromised.  In this paper we present an empirical decontamination technique for extracting superior, decontaminated spectra of transient objects from UVOT grism images.  Our empirical decontamination technique parallels image subtraction techniques commonly used in photometry.  We present a description of our empirical decontamination technique and demonstrate its accuracy in extracting flux calibrated spectra by comparison with long-slit spectra of SN 2011fe, and demonstrate its practical application with spectra of iPTF14bdn.  In the Appendix we present UV spectra of several additional SNe Ia extracted using our technique.

\section{UVOT Background}
\subsection{Observing Modes}
Both grisms have the ability to observe in either nominal or clocked mode.  Clocking is an observing mode in which the filter wheel containing the grisms is positioned to partially obscure the light path such that the diffracted spectrum of interest falls upon a region of the detector that is not exposed to sky.  Using this procedure $\sim \frac{2}{3}$ of a typical spectrum's wavelength range falls upon the obscured portion of the detector while the remaining $\sim \frac{1}{3}$ falls upon a portion of detector exposed to sky.  For the UV-grism the region of the spectrum which falls on this sky-illuminated region is the shortest wavelength region ($1700-3000$\,\AA.)  Clocked mode is most often used because it prevents projecting the redder portion of the diffracted spectrum of interest on top of nearby contaminant sources, and it provides a lower background and thus smaller coincidence loss correction \citep{Kuin2015}.  

\subsection{Contamination in Grism Images} \label{contaminationsection}
Contamination can occur in two ways: where the dispersed target spectrum lays $1)$ on the same pixels as light from a different source, or $2)$ on pixels very nearby a different light source.  In the first case, the target spectrum is overlaid on pixels simultaneously measuring light of either a zeroth or first order image of the contaminating light source.  Here, the target photons are indistinguishable from the contaminating photons such that the two cannot be disentangled.  When this occurs no reliable flux measurements of the target can be made at the contaminated wavelengths.  In the second case the photons from the target and contaminant sources do not fall on the same pixels.  Instead, the contaminant photons fall in the regions of the detector used for calculating the background flux.  This influences the measured background value in the image processing phase and prevents an accurate determination of the background flux for a range of wavelengths.  Both forms of contamination are problems for targets near diffuse and clumpy objects, such as galaxies.

Contamination can sometimes be averted in the planning stage by simulating the dispersion orientation relative to contaminant sources prior to observation using the {\it simgrism} simulation \footnote{http://idlastro.gsfc.nasa.gov/ftp/landsman/simgrism/simgrism.pro}.  This program allows an observer to determine a spacecraft roll angle that provides a minimal amount of overlap between the diffracted spectrum and bright sources in the Digitized Sky Survey\footnote{Digitized Sky Survey data was obtained from the Mikulski Archive for Space Telescopes (MAST). STScI is operated by the Association of Universities for Research in Astronomy, Inc., under NASA contract NAS5-26555. Support for MAST for non-HST data is provided by the NASA Office of Space Science via grant NNX09AF08G and by other grants and contracts.}.  During observation, use of the clocked mode reduces zeroth order contamination of the spectrum of interest, but unfortunately does not prevent contamination of the UV portion of the spectrum.  In the image processing phase the \uppercase{uvotpy} \citep{Kuin2014} software is designed to minimize the effects of contamination as much as possible.  In calculating the background flux along the dispersion direction \uppercase{uvotpy} allows the replacement of bright sources inside the background measurement channels with averaged values.  This processing method is very effective at removing contamination caused by point sources whose zeroth or higher order spectra fall within the background measurement regions. However, it does not provide any reliable solutions for accessing spectra which fall on top of a zeroth order source, which lie near clumpy or diffuse sources, or which lie in a field dominated by strong sources.  

\section{Method of Decontamination}
\subsection{Motivation}
Since shortly after its launch in 2004, {\it Swift} UVOT has been photometrically and spectroscopically observing supernovae.  To date $\sim 30$ SNe Ia and $\sim 20$ core collapse SNe have been targeted for grism observations.  Despite the large number of targets, the number of published UVOT spectra of SNe remains quite small \citep{Bufano2009, Foley2012, Bayless2013, Brown2014, Brown2015, Margutti2014}.  This is partly due to the fact that roughly half of the SNe Ia grism observations contain contamination which cannot be removed using the normal image processing methods and thus render the final spectral features and calibrations suspect or useless.  Because SNe Ia are transient objects, the contaminated data cannot be re-observed and is thus lost if the contaminating background sources cannot be removed.  Because of the high value of space-based UV spectral observations, we opted to develop a method of removing as much of the contamination as possible.  To do this and make more of the sample of grism observations of SNe available for analysis we have developed an empirical decontamination technique for {\it Swift} UVOT grism images.  The process is akin to image subtraction for direct images and thus requires that any sources of contamination remain unchanged between data and template image acquisitions.  

While designed for use with SNe Ia, this method should be equally applicable to any transient flux source that is sufficiently faint after outburst or any object which moves over time on an unchanging patch of sky.  For example, the UVOT grisms have been used in the study of active galactic nuclei \citep{Cackett2015} and comets \citep{Bodewits2011}.  Since the method was developed for the purpose of recovering SNe spectra we will use SNe as the examples for the remainder of this paper.

Templates of $17$ SNe Ia are being observed as part of the {\it Swift} Cycle 11 Guest Investigator Program \#1114103: "Decontaminating the {\it Swift} UV-Grism Sample of SNe Ia to Measure the UV Diversity" (PI: Suntzeff, Science PI: Smitka).  The data gathered as a part of this observing program will be analyzed using the method described herein and the final decontaminated spectra made public in an archive upon completion.

The method consists of four main steps: $1)$ template observation, $2)$ template registration, $3)$ flux scaling, and $4)$ spectral extraction. 

\subsection{Template Observations} \label{templatesection}
Our empirical decontamination technique requires that UVOT observe template images which reproduce the original data observations as closely as possible.  A template image is a grism image observed after the target flux has sufficiently faded.  We have determined that the following critera should be adhered to in order to ensure that template observations can be successfully implemented in the empirical decontamination process: 
\begin{itemize}
\item{{\it Flux:} The template images must be observed after the target flux has become fainter than the UVOT detectability limit for the entirety of the grism's wavelength range.  This ensures that the template images contain only flux from possible contaminant sources and not residual flux from the target itself.  For most supernovae including SNe Ia, fulfilling this requirement requires that templates be observed at least one year after peak brightness.}  
\item{{\it Pointing:}  The spacecraft should be positioned to reproduce the right ascension (RA) and declination (dec) of the data observations as closely as possible.  In general, the data exposures exhibit offsets of a less than an arcminute on the sky.  The template observations should be designed such that a small translation of the template detector image (*dt.img) can be applied to force an overlay of the template image upon each of the data detector images.  The UVOT effective area is known to vary as a function of location on the detector, so an initial 'slew-in-place' pointing should be utilized to reproduce the original pointing\footnote{A 'slew-in-place' pointing consists of a normal slew command followed by a second identical slew command once the initial slew has completed.  This maneuver provides a higher pointing accuracy than a single slew command.}.  This will ensure that the photon light path through the telescope optics are reproduced as accurately as possible and the effect of the spatially-varying flux sensitivity is minimized.}
\item{{\it Roll Angle:} The spacecraft should be positioned to reproduce the roll angle of the data observations as closely as possible.  For typical observations the roll angle of the spacecraft is set as an integer number of degrees \footnote{Early in the {\it Swift} mission fractional degrees were used.}; a template image is only applicable to data images of the same integer roll angle.  Due to spacecraft orientation constraints, special care must be taken in the planning of the template observations.  A template observation must be observed within about two weeks of the calendar date of the data observations to ensure that the RA, dec and roll angle can be reproduced.  This constraint is in place due to the limited ability of the spacecraft to maneuver into an orientation which reproduces the RA, dec and roll angle of the data observations.  The windows of availability repeat each year due to the cyclic nature of orbital mechanics.}
\item{{\it Integration Time:} The total exposure time of a template observation should be comparable to the total exposure time of the data observation.  This ensures a similar signal-to-noise factor in both images.  Individual template observation exposures can be coadded to obtain the required integration times if necessary.}
\item{{\it Clocking:}  Template images must be gathered in the same grism mode as the corresponding data images.  For example, clocked data images must be decontaminated using clocked template images.}
\end{itemize}

\subsection{Template Registration}
The next step in the empirical decontamination technique is registering the template image to overlay the data images.  This process is necessary because the pointing differences between exposures can vary by up to one arcminute ($\sim 100$~pixels).  As a result, this process must be carried out individually for each exposure (image extension).  A simple translation of the template detector image is all that is necessary.  Translations are typically 100 pixels or fewer in each dimension when the guidelines of Section \ref{templatesection} are followed.  Translations greater than this amount should be avoided due to the spatially varying sensitivity of the detector.  The flux calibration of \citet{Kuin2015} accounts for the overall spatial variation of the detector, but not for the difference in sensitivity resulting from the data and template images being slightly offset.  For clocked UV-grism template images with an offset of $100$\, pixels or fewer we estimate that the uncertainty in measured flux introduced by the registration process is  $3\%$\, or less.

For the purpose of registering the images we were unable to find any existing software packages capable of operating on grism images.  Manually translating the templates for each data exposure is possible, but is very time consuming.  To reduce the burden on the user we have developed an automated translation algorithm which behaves as follows: $1)$ The user inputs a crude estimate of the offset between the two images.  This is done by supplying the x and y coordinates of a zeroth order feature common to both images.  $2)$ A $20\times20$ grid is defined with axes corresponding to integer value pixel shifts in the x and y plane ranging between the estimated offset $\pm10$ pixels.    
We investigated the utility of registering the images by fractional pixel shifts and found no improvement in registration quality.  So, integer pixel shifts are favored for simplicity and computational efficiency.  $3)$ The template image is translated in the x and y plane by values corresponding to each point on the grid.  For each translation the template image is subtracted from the SN data image and the standard deviation of the pixel values on the entire residual image is calculated.  The grid is iteratively populated with the standard deviation values in this way.  $4)$  The grid point with the minimum standard deviation value is identified as the best translation.  $5)$  The template image is translated by the optimum value.  $6)$ A subtracted residual image is provided to the user as a visual quality check of the alignment.  We have tested this method and found it to provide good results for pixel shifts of up to $\sim 100$\, pixels in each dimension.  Templates requiring greater pixel shifts than this are not available to us.  We do not advise applying this method for pixel shifts greater than this due to the spatially varying sensitivity of the detector, as previously discussed.

\subsection{Flux Scaling}
Next, the registered template image must be flux scaled to account for differences in the integration times of the data images and the template image.  The purpose of this step is to scale the flux on the template image such that sources of constant brightness have equivalent counts in both data image and the scaled template image.  This step can be avoided if the integration time of both the data and template images are identical.  A simple multiplicative scaling of the counts on the registered template detector image is all that is required.  The scaling factor is calculated as the ratio of the exposure times of the template and data images.  

In some cases an additional small flux scaling factor is necessary to account for the $\sim 1$\% per year sensitivity loss of the UVOT detector \citep{Breeveld2011}.  This is typically required in cases where a sufficiently long span of time elapses between the data and template observations that the UVOT sensitivity changed significantly.  In these cases, the correction accounts for a slight faintness of the template image relative to the data image.  In practice, this effect can be corrected $1)$~ theoretically by assuming a value for sensitivity loss as a function of time and multiplying the template image by the factor and length of time elapsed between observations, or $2)$~ empirically by measuring fluxes of constant brightness sources in both images, calculating a flux ratio of the two and multiplying this factor into the template image.  

\subsection{\uppercase{uvotpy} Extraction}
The \uppercase{uvotpy} software for extracting, flux calibrating and wavelength calibrating UVOT grim spectra includes an option to supply an external background image, yet does so in a way that accurately calculates coincidence loss for the UVOT detector \footnote{The \uppercase{uvotpy} software is available for download at http://www.github.com/PaulKuin/uvotpy}.  The change from the typical extraction process lies in how the measurement of the background is performed.  In a typical \uppercase{uvotpy} extraction the background flux imposed on the target spectrum is not directly measured, but is estimated from the flux contained in channels lying parallel to the dispersion direction directly above and below the channel containing the target spectrum.  The background flux along the dispersion direction is then calculated from the sigma clipped mean to remove bright sources lying within the channels.  Next, the smoothed flux within these channels and the coincidence loss is calculated using the flux within the extraction channel plus the background estimate.  The corrected background is then subtracted from the total flux to obtain the coincidence loss corrected target flux.  When the empirical decontamination technique is used, the contaminant flux for each pixel is measured directly from the corresponding location on the registered and scaled template image.  The coincidence loss is then calculated using the total flux from the extraction channel on the data image and for the background measurement from the extraction channel on the template image.  The flux errors are based on the total flux in the target spectrum and that from the template image.  The primary benefit of this process is that it decouples the flux of the target object from any underlying sources within the extraction channel while properly correcting for the effect of coincidence loss on the detector due to both target and contaminant photons.  

The extraction process described above is necessary due to the photon counting nature of the UVOT detector and the non-linear nature of the coincidence loss correction.  The UVOT detector does not directly measure incident photons, but rather incident photons interact with a photocathode which produces a cascade of photons that is recorded as a splash on the detector.  A centroiding process of the splash then calculates the spatial location of the incident photon.  Coincidence loss occurs for bright sources as a result of the UVOT detector's finite frame time while recording these splashes.  When multiple photons are incident upon the photocathode during an individual readout frame the multiple splashes can be recorded as a single splash if the incident spatial locations are near each other.  In this case photons remain uncounted.  If a correction is not applied the corresponding flux measurement will underestimate the true flux of the source.  The coincidence loss correction for all UVOT grism spectra is performed by the UVOTPY extraction software and is documented in detail in \citet{Kuin2015}. 

A classical image subtraction process in which the registered and scaled template image is subtracted from the SN data image would succeed in decoupling the target and contaminant fluxes, but would subsequently be improperly flux calibrated due to neglecting a proper treatment of the coincidence loss occurring on the detector as it counts photons from both sources simultaneously.  This being said, classical image subtraction is useful for applications which do not directly lead to a calibrated flux measurement.  For example, we implement subtracted images as quality checks in the registration and flux scaling processes.  

\section{Application}
To demonstrate the accuracy of the flux calibration of our empirical decontamination technique we present a comparison with well-calibrated long-slit spectroscopy of SN 2011fe near peak brightness.  We also present an analysis of iPTF14bdn to demonstrate our use of the empirical decontamination technique to remove contaminant features that are easily confused with SNe features.  Additional SNe Ia spectra extracted using our technique are available in the Appendix.

\subsection{SN 2011fe} \label{section2011fe}
Supernova 2011fe data affords us the opportunity to demonstrate the accuracy of the flux recovery from the SN using our empirical decontamination technique.  Due to its location in the nearby galaxy M101, SN 2011fe was a very bright SNe Ia.  For this reason it was heavily monitored by several observatories, including both the {\it Hubble Space Telescope} ({\it HST}) Space Telescope Imaging Spectrograph (STIS) \citep{Mazzali2014} and {\it Swift} UVOT \citep{Brown2012}.  The UV slit spectra of SN 2011fe obtained using STIS are superior to those of UVOT because {\it HST} has a much larger aperture, the high spatial resolution and narrow slit produce spectra free of contamination and have an absolute flux calibration accurate to $5$\%~  \citep{Bostroem2011} and can thus be used as benchmarks for comparison.  

\begin{figure*}[th]
\centering
\subfigure{%
\includegraphics[height=2.0in,width=2.0in]{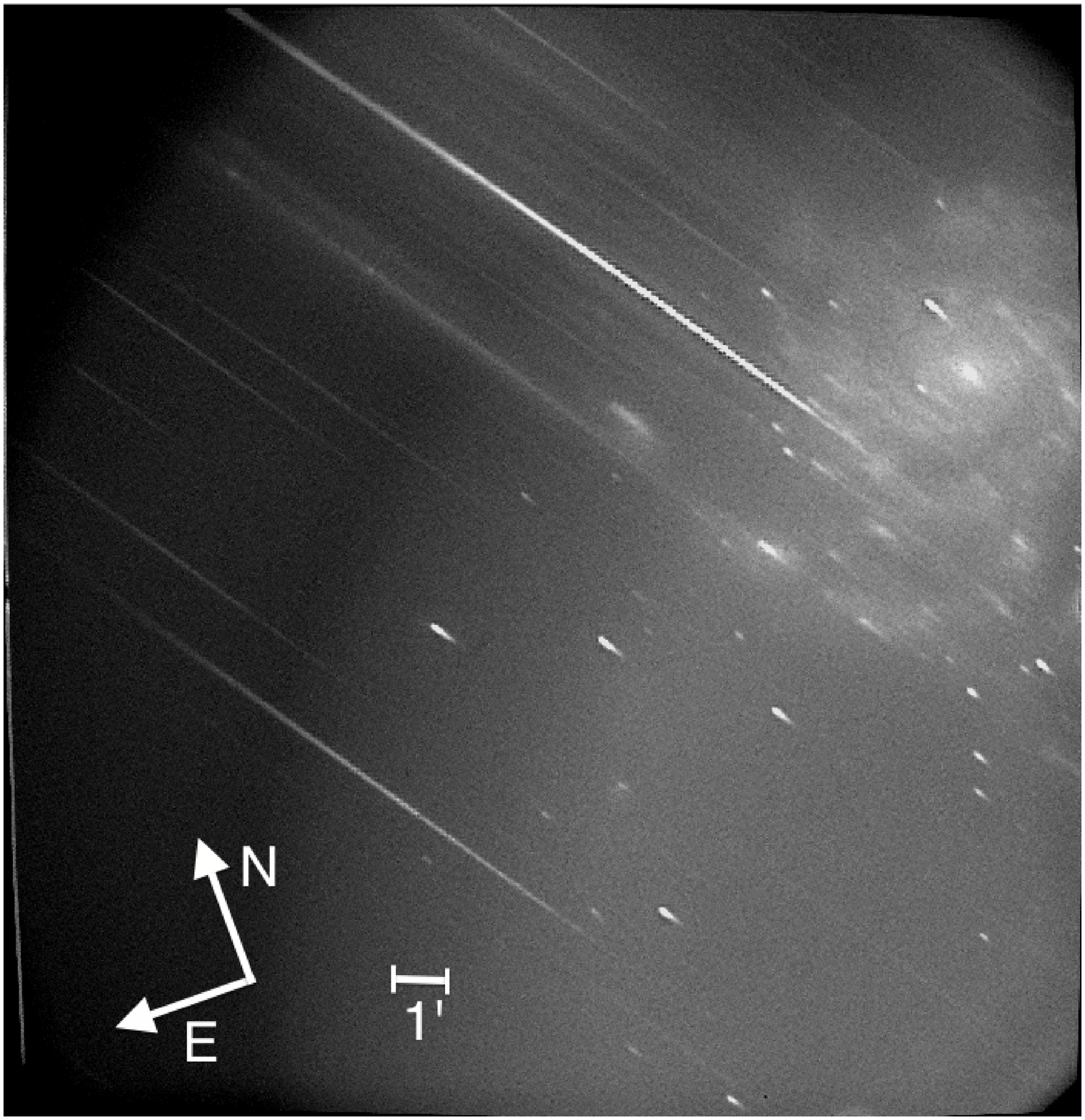}
\label{2011feimage1} }
\subfigure{%
\includegraphics[height=2.0in,width=2.0in]{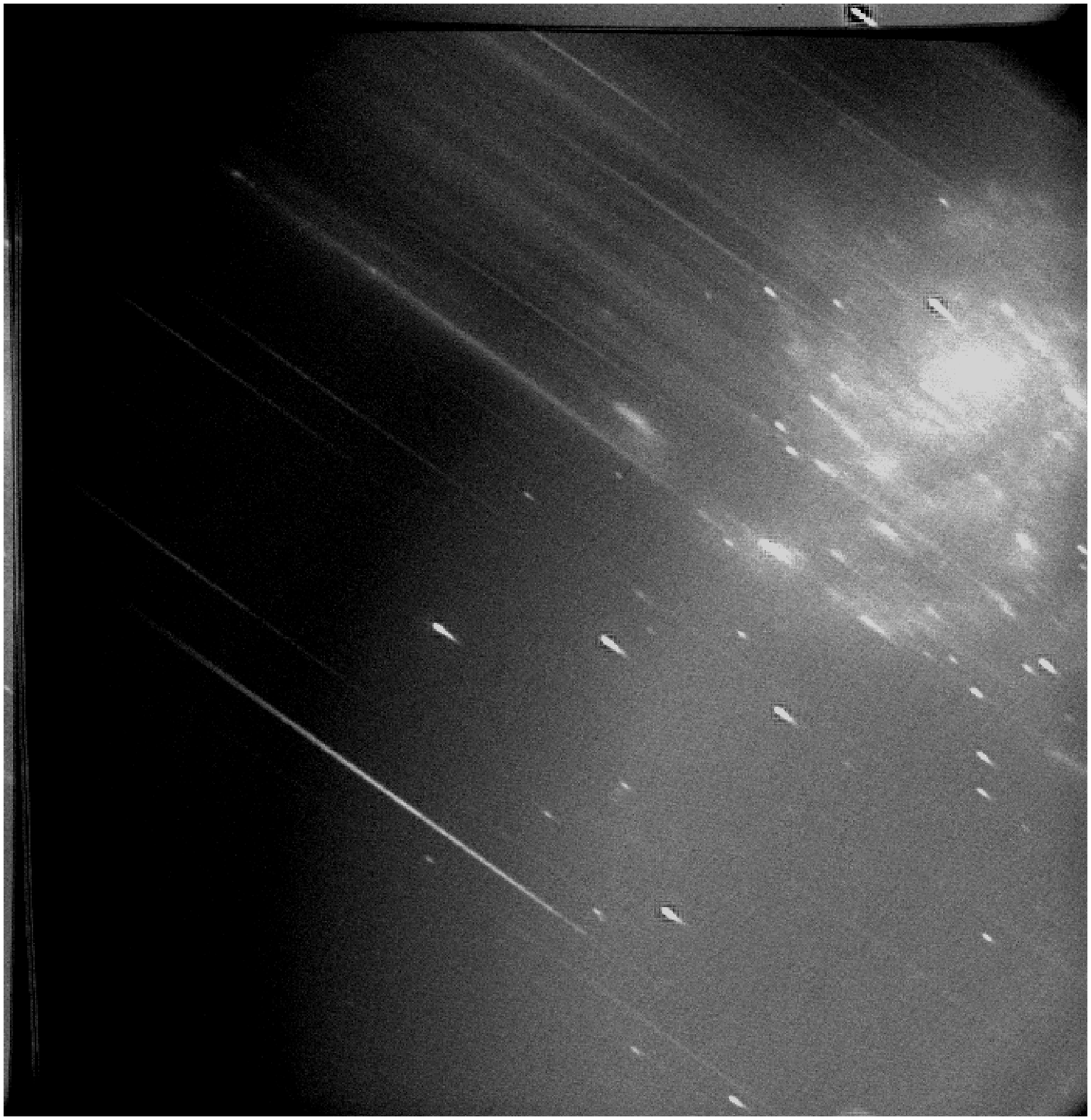}
\label{2011feimage2} }
\subfigure{%
\includegraphics[height=2.0in,width=2.0in]{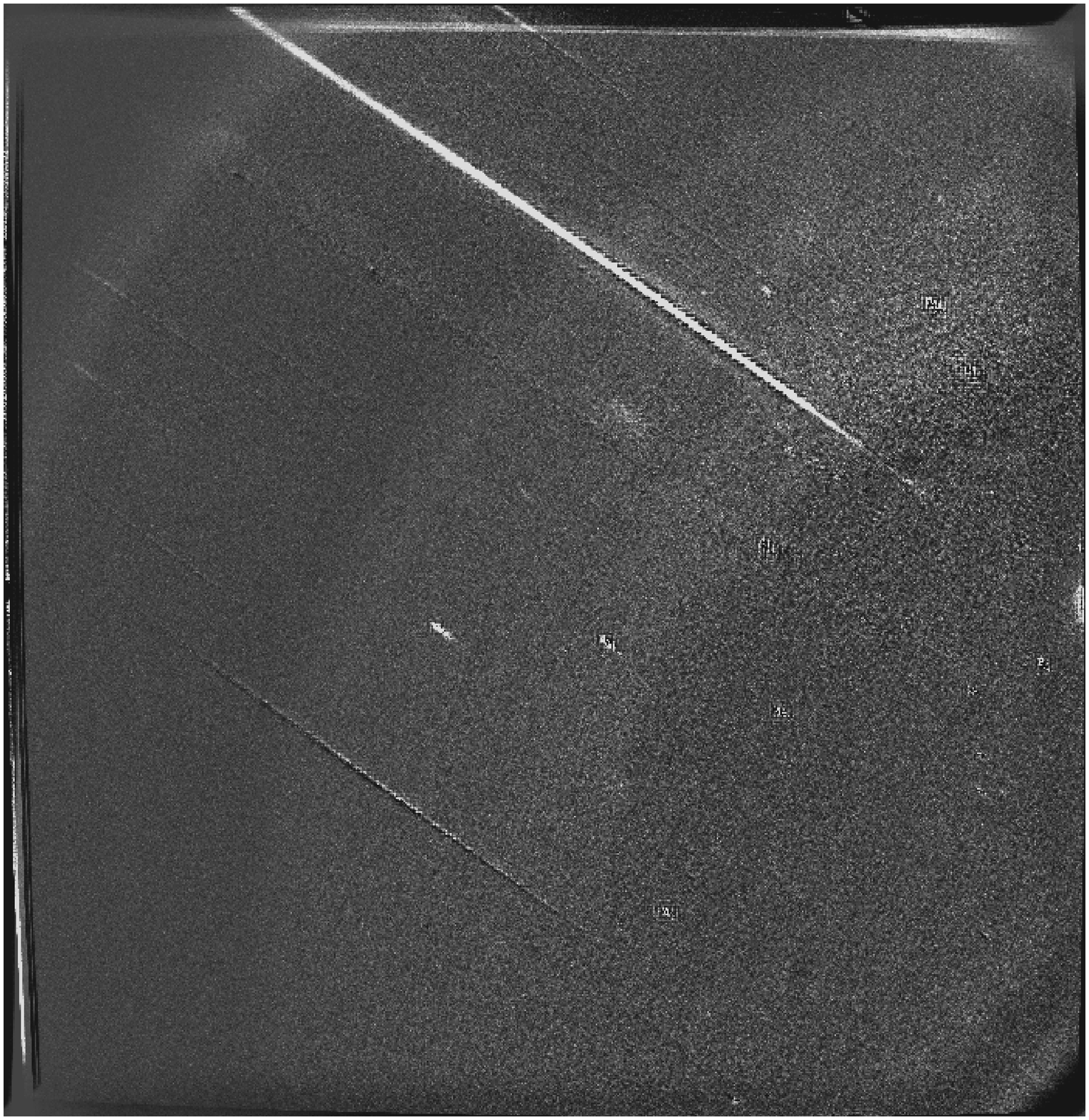}
\label{2011feimage3} }
\caption{{\it Swift} UVOT U-grism images.  {\it Left:} A data image of SN 2011fe.  The zeroth order image of SN 2011fe lies just outside of the frame to the right, the first order diffracted spectrum is the brightest feature and appears partially overlaid on the host galaxy M101.  Wavelength of the diffracted spectrum increases from bottom right to upper left.  {\it Center:} A registered and flux-scaled template image for SN 2011fe.  This image was taken two years after the SN data images.  {\it Right:}  The residual of subtracting the registered and flux-scaled template image from the SN data image.  The SN light is decoupled from the galaxy light.}
\label{2011feimages}
\end{figure*}

The UVOT spectra of SN 2011fe exhibit more contamination than any other SN observed with UVOT.  As a result of the proximity of the host galaxy, the location of the SN within the galaxy, and {\it Swift}'s roll angle availability, the grism images exhibit both types of contamination described in Section \ref{contaminationsection}.  In all observations, the first order spectra are projected on top of resolved zeroth order and dispersed galaxy features and the background measurement channels near the diffracted spectra contain both diffracted and zeroth order light from clumpy sources (Figure \ref{2011feimages}.)  Using normal \uppercase{UVOTPY} extraction procedures the SN and host galaxy fluxes cannot be decoupled and all SN spectral information blueward of 2500\,\AA~ is contaminated.  In this region one does not know a priori which extracted spectral features are from the SN and which are from the galaxy.  However, SN 2011fe template images contain only the underlying galactic flux and the STIS spectra contain only the SN flux, so in this case we can test the ability of our empirical decontamination technique to decouple the SN and host galaxy fluxes while assessing the accuracy of the recovered SN flux.

\begin{figure}[h]
\plotone{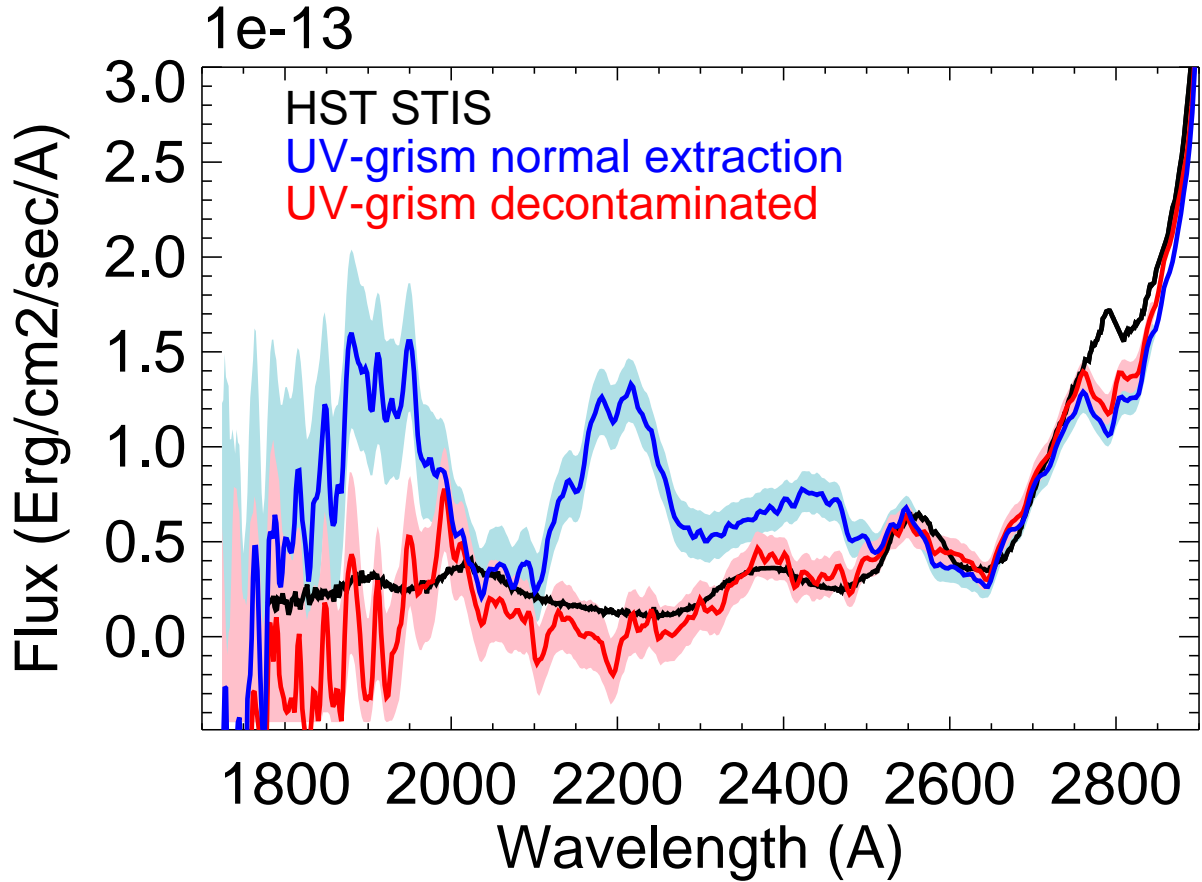} 
\caption{{\it Swift} UVOT U-grism spectra of SN 2011fe extracted using the normal \uppercase{uvotpy} method and our empirical decontamination technique are compared to {\it HST} STIS spectra.  The UVOT and STIS spectra were gathered just hours apart while the SN was very near maximum brightness.  The contamination to the UV portion of the UVOT spectrum is removed and the agreement with the STIS spectrum is improved.  The noise at the shortest wavelengths is due to a combination of low photon rates and decreased detector sensitivity.}
\label{2011fespec}
\end{figure}

We carried out our empirical decontamination technique using the clocked UVOT UV-grism data of SN 2011fe observed on 2011 September 10 (obsid: 00032094004) and the template we observed as a {\it Swift} target of opportunity two years after the initial observations on 2013 September 11 (obsid: 00032094018).   For the template images the pointing of the spacecraft was recreated to within $30$\, arcsec in both RA and dec and within $4.5$\,arcmin in roll angle.  At the time of the original observation the SN was at a phase of $0.11$~days past B-band maximum light.  We chose this epoch for analysis because simultaneous STIS spectra exist and provide us with a benchmark for comparison.  Figure \ref{2011fespec} shows that we were able to decouple the SN from the contaminant fluxes and demonstrate agreement between the calibrated UVOT and STIS SN fluxes as blueward as $2000$\,\AA.  After subtraction of the host galaxy flux, there is little flux remaining from the supernova shortward of $2500$\,\AA.  In the absence of HST/STIS or subtracted UVOT spectra, the strong features in the unsubtracted spectra could have been misinterpreted as coming from the SN and therefore resulted in unnecessary (and incorrect) theoretical interpretations.

\subsection{iPTF14bdn}
Supernova iPTF14bdn \citep{Cao2014} provides us with an example of how our empirical decontamination technique is beneficial to the scientific analysis of UVOT grism spectra when they are the only available spectra.  This SN was the first 1999aa-like SN Ia to have an UV spectral series published \citep{Smitka2015}.  Prior to using this technique, upon comparison to normal SNe Ia the pre-maximum light spectra of iPTF14bdn displayed much greater UV flux in two regions of the spectra: between $2800-3200$\,\AA~ and at $2500$\,\AA.  The $2800-3200$\,\AA~ feature was of the same shape as normal SNe Ia features, but the relative strength of the feature compared to optical features was much stronger.  The $2500$\,\AA~ feature was anomalous and very unlike any UV spectral feature seen in any other SNe Ia.  This feature was slightly conspicuous because it appeared equally strong in spectra only associated with a spacecraft roll angle of $316$~degrees, and could therefore possibly be due to a faint zeroth order contaminating source lying inside the extraction channel at this orientation.  Alternatively, because the feature appeared only in the earliest observations and not the later ones it was possibly an evolutionary feature of the SN which had never been observed before.  While both of these features are very interesting, the $2500$\,\AA~ feature was more scientifically pressing due to its uniqueness and the potential for new physics it promised.  

\begin{figure}[h]
\plotone{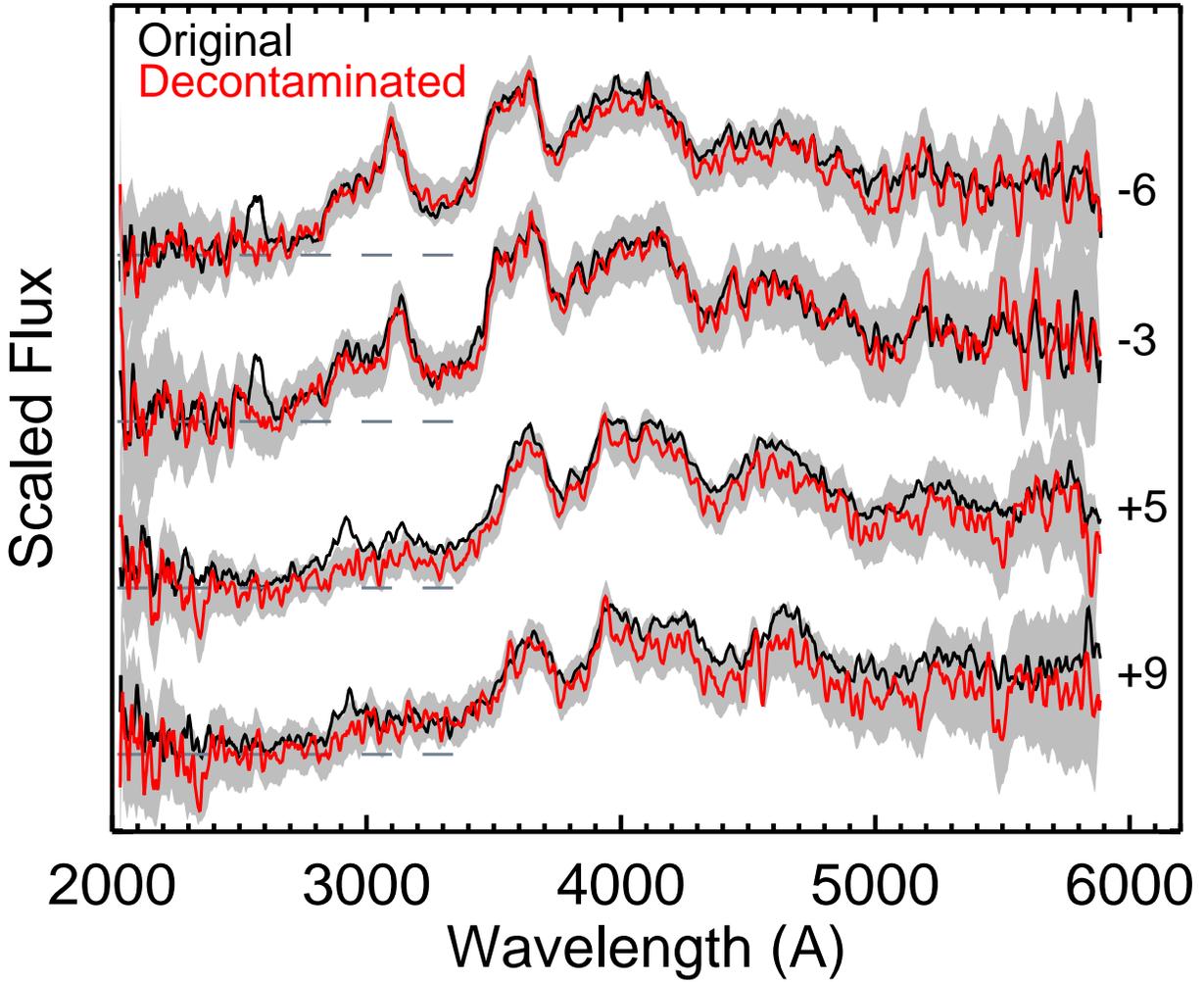} 
\caption{{\it Swift} UVOT spectra of SN iPTF14bdn extracted using the normal \uppercase{uvotpy} method (black) and our empirical decontamination technique (red).  The fluxes shown have been normalized.  Phases shown are in units of days since B-band maximum light.  The flux spikes near $2500$\AA~ in the two pre-maximum light spectra are removed by our decontamination process.}
\label{14bdnspec}
\end{figure}

To determine whether these spectral features were associated with the SN or contamination we carried out template observations and performed the empirical decontamination technique on the SN data images.  We observed the templates one year after the SN data observations as part of our {\it Swift} Guest Investigator project.  For the template images the original pointing of the spacecraft was recreated to within less than one arcsecond in RA and dec and $17$\, arcsecond of roll angle.  We found no residual SN signal in the template images.  Details of the SN and template observations are presented in Table \ref{14bdntable}.  The original and decontaminated spectra are compared in Figure \ref{14bdnspec}.  It is shown that the empirical decontamination technique removed the flux spike at $2500$\,\AA~ without significantly modifying the feature at $2800-3200$\,\AA~in the pre-maximum spectra.  This enabled us to conclude that the $2500$\,\AA~ feature was a contaminating source and that the $2800-3200$\,\AA~ feature was associated with the SN and worthy of further analysis as described in \citet{Smitka2015}.  

\section{Treatment of Errors}

The UVOT is a photon counting instrument, where each imaging observation is made up of a finite number of frames.  This means that the errors follow a Poison distribution as long as the number of counts per pixel in an observation is much smaller than the number of frames.  In cases where this is not so, the errors need to be computed according to the formalism presented in \citet{Kuin2008}.  

The treatment of errors is implemented within the \uppercase{UVOTPY} software.  Since the observed spectrum is determined by subtracting a background, the total error in the observed spectrum is the root mean square sum of the errors in background and spectrum.  The same procedure is employed for both the default background produced in UVOTPY (described in Kuin et al. 2015) and for the background supplied from matching a late-time observation of the field in its place. 

In the case of spectral extraction, \uppercase{UVOTPY} resamples the spectrum while rotating the image for extractions (see \citet{Kuin2015}).  As a result, a correlation is introduced between the errors in neighboring spectral bins.  Tests show that assuming three bins are correlated will bring the values of the $\chi$-squared test to the expected values \citep{Kuin2009}.  

The empirical decontamination technique works similarly as when using the normal \uppercase{UVOTPY} method in which the background is estimated by sigma clipping and smoothing the original image.  For the following discussion we will assume that the changes in the field are only due to the fading away of the transient object spectrum.  The normal \uppercase{UVOTPY} background, being an extrapolation from nearby regions, is almost certainly less accurate because the transient spectrum may lie in pixels also containing light from weak sources.  This includes either zeroth or first order light, which cannot be removed (see the spectra of SN 2010ev and SN 2012cg in the Appendix for examples.)  Also, the opposite may be true where the spectrum lies over a slightly less crowded part of the background (see the spectra of SN 2005df in the Appendix for an example.)  The estimation of the normal \uppercase{UVOTPY} background also becomes less accurate when the source is near a number of strong first order spectra.  These are all systematic errors above the Poisson errors mentioned already.  The method of this paper does not suffer from these kinds of systematic errors.  The gain is mainly obtained where the spectrum is up to $10$\, sigma above the background noise, but also in removing weak underlying zeroth and first order contaminations.  The contaminating source count rate raises the count rates in the supplied background and the errors are correctly treated for the higher contribution from the contaminating source. Surprisingly, in the \uppercase{UVOTPY} spectral extraction the same contaminating source count rate will be added to the source spectrum and add to the random error there (as well as being an unidentified systematic error), so the difference in the size of the random error between the two methods is small.  However, the difference and advantage lies in removing the systematic errors by using the empirical decontamination technique.  

Coincidence loss, has a spatial component that affects the spectral profile when very large.  The reason lies in the photon splash centroiding which is based on three physical detector pixels - equal to 24 (sub)pixels \citep{Roming_etal_2005} - and the coincidence loss affects the centroiding.   As a result, for very bright sources, a zeroth order field star near the target spectrum will have a large error or become unrecoverable (see the discussion of SN 2012dn in the Appendix for an example.)  When extremely strong contamination is present both the normal \uppercase{UVOTPY} extraction and the empirical decontamination technique cannot recover the affected portions of the spectrum.  In the UVOT grism calibration documentation, \citet{Kuin2015} determined that the coincidence loss correction is found as a best fit to a set of calibration observations.  For large count rates, where the total count rate in the spectrum needs to be regarded, the error increases.  A limiting count rate equivalent to $0.97$\, counts per bin per frame was selected as an upper limit as above that rate the error in the coincidence loss correction became much larger than 20\%.  For a more detailed description of the coincidence loss correction the reader is referred to \citet{Kuin2015}, section 7.

\section{TRUVOT Software}\label{software}
We have designed the TRUVOT software pipeline \citep{Smitka2015b} to perform our empirical decontamination technique.  The pipeline consists of IDL and \uppercase{uvotpy} Python routines \footnote{The IDL portions of the software pipeline can be run under the IDL virtual machine without an IDL license.}.  This software and a working example of the SN 2011fe empirical decontamination extraction of Section \ref{section2011fe} is available at http://github.com/mikesmitka/truvot.  The software is also available in the \uppercase{UVOTPY} distribution.

\section{Conclusion} \label{conclusion}
We have presented a new technique for decontaminating {\it Swift} UVOT grism spectra of transient objects.  We described the template image requirements and image processing steps necessary to successfully implement the technique.  Two examples were given to demonstrate the accuracy of the flux and wavelength calibrations, and the scientific applicability of the technique.  The software which implements this technique was briefly discussed and has been made available to the community.

\acknowledgements
The authors would like to thank the {\it Swift} science operation team, Daniele Malesani in particular, for their assistance with observing the template images of our {\it Swift} Guest Investigator project.  M. Smitka is supported by the {\it Swift} Guest Investigator Program through grant NNX15AR51G.  P. J. Brown and the Swift Optical/Ultraviolet Supernova Archive are supported by NASA's Astrophysics Data Analysis Program through grant NNX13AF35G.  We acknowledge the use of public data from the Swift data archive.  We acknowledge funding from the UK Space Agency for work performed at MSSL/UCL.

\appendix \label{appendix}
\begin{center}
      {\bf APPENDIX}
\end{center}
Additional SNe Ia spectra with templates observed as part of our {\it Swift} Guest Investigator project are presented here.  Details of the spectral observations and reductions can be found in Table \ref{appendixtable}.

\subsection{SN 2005df}
\begin{figure}[h]
\plotone{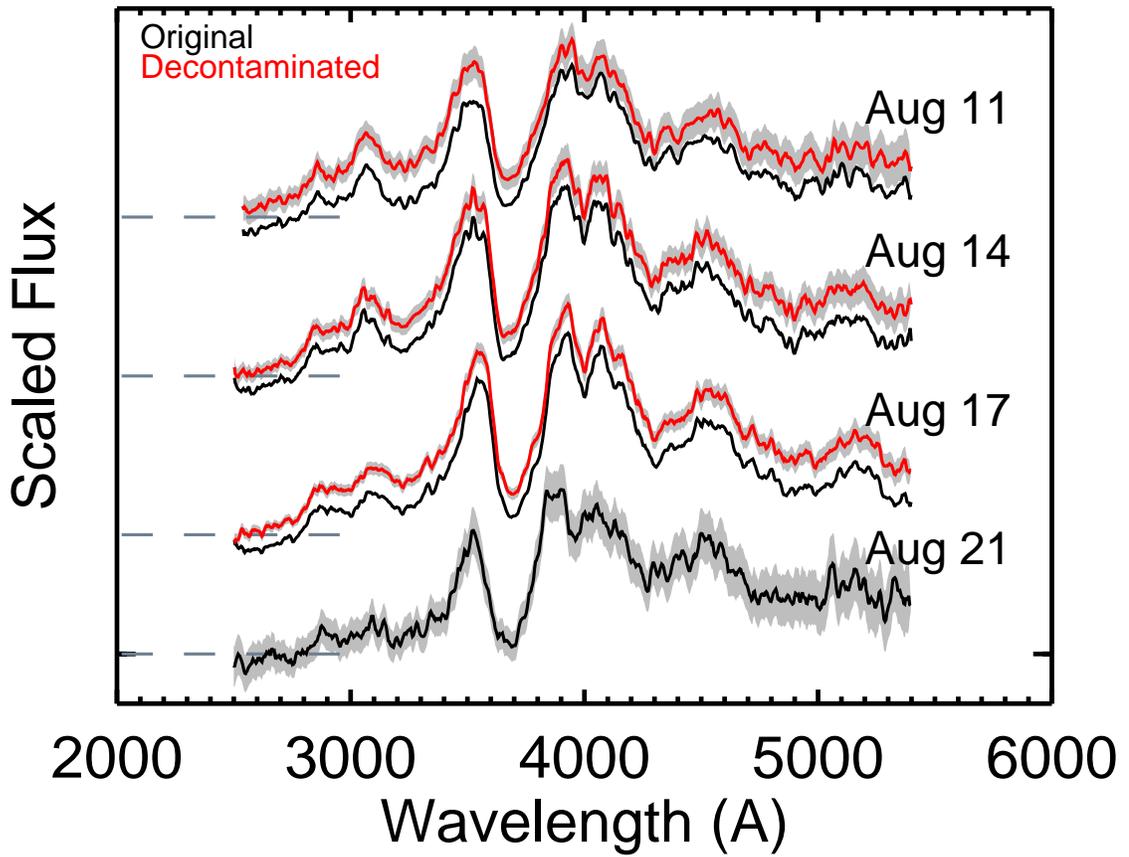} 
\caption{Spectra of SN 2005df extracted using the regular UVOTPY extraction and our empirical decontamination technique.  The offset between the spectra is due to the first order dispersed host galaxy light appearing in the background estimation region of the regular extraction and not appearing within the dispersed SN light measurement region.}
\label{2005dfspec}
\end{figure}

Four spectra of SN 2005df extracted using our empirical decontamination technique are presented in Figure \ref{2005dfspec}.  An earlier reduction of these observations were presented by \citet{Bufano2009}.  We were unable to remove the severe contamination from a saturated zeroth order field source below $2500$\,\AA~  and obtained no usable spectra below this limit.  The spectra of \citet{Bufano2009} predate the \uppercase{uvotpy} software, and so the spectra presented here are favorable because they have been reduced using the most recent calibrations and updated software.  We calculated phases relative to B-band maximum light using the peak value of \citet{Milne2010} occurring at $JD\,=\,2453598.825$.  

\subsection{SN 2009dc}
\begin{figure}[h]
\plotone{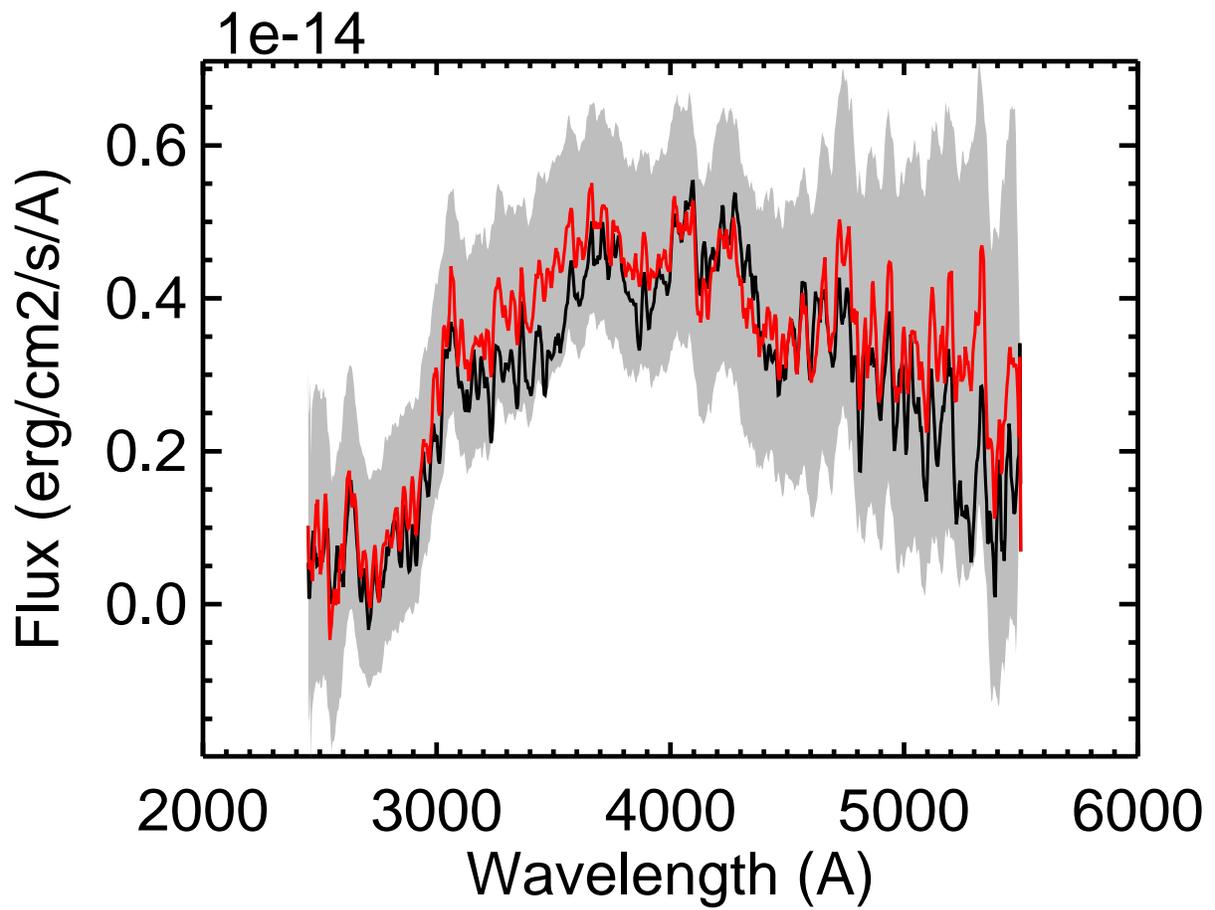} 
\caption{A spectrum of SN 2009dc from 2009 May 01 extracted using the regular UVOTPY extraction and our empirical decontamination technique.  This spectrum was originally presented by \citet{Brown2014}.}
\label{2009dcspec}
\end{figure}

A UVOT spectrum of SN 2009dc was presented in \citet{Brown2014}.  We re-extracted this spectrum using our empirical decontamination technique and present it in Figure \ref{2009dcspec}.  The spectral features blueward of $2700$\,\AA~ are still present following decontamination.  The phase of this observation was calculated using the UVOT photometry B-band time of maximum light of $JD\,=\,2454947.3$ \citep{Brown2014}.

\subsection{SN 2009ig}
\begin{figure}[h]
\plotone{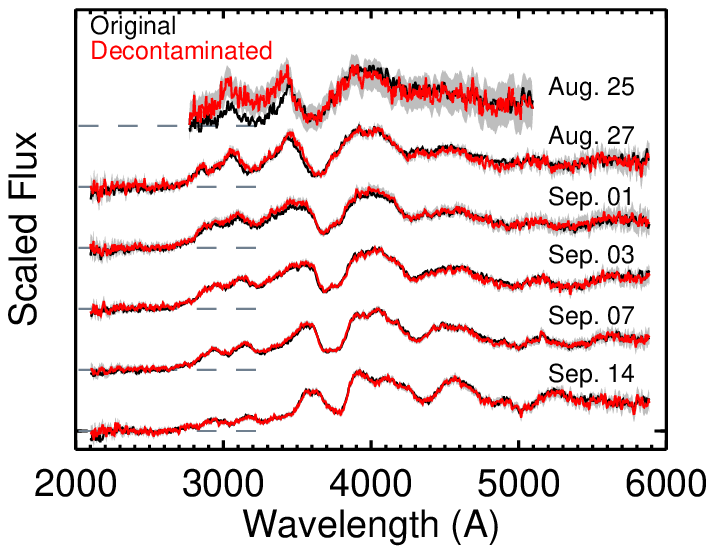} 
\caption{Spectra of SN 2009ig extracted using the regular UVOTPY extraction and our empirical decontamination technique.}
\label{2009igspec}
\end{figure}

Six spectra of SN 2009ig extracted using our empirical decontamination technique are presented in Figure \ref{2009igspec}.  We calculated phases relative to B-band maximum light using the peak time of \citet{Foley2012} occurring at $JD\,=\,2455080.5$. 

\subsection{SN 2010ev}
\begin{figure}[h]
\plotone{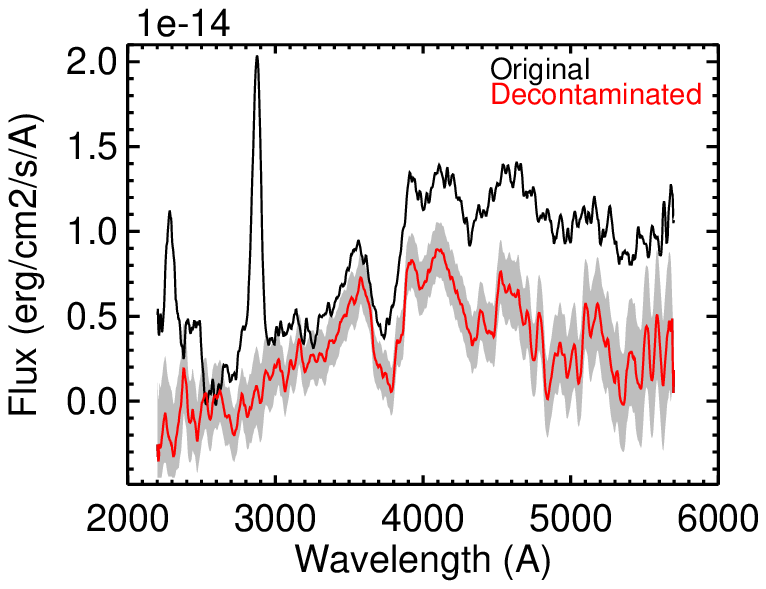} 
\caption{Spectra of SN 2010ev from 2010 July 05 extracted using the regular UVOTPY extraction and our empirical decontamination technique.  The offset between the spectra is due to the first order dispersed host galaxy light being superimposed on the dispersed SN spectrum and not appearing within the background estimation region.}
\label{2010evspec}
\end{figure}

A spectrum of SN 2010ev extracted using our empirical decontamination technique are presented in Figure \ref{2010evspec}.  We attempted to decontaminate an additional observation from 2010 July 12 but were unable to extract a viable spectrum due to contamination from an extremely bright nearby bright zeroth order source for which a reliable coincidence loss could not be calculated.  The phase of this observation was calculated using the UVOT photometry B-band time of maximum light of $JD\,=\,2455385.4$ from the Swift Optical/Ultraviolet Supernova Archive (SOUSA) \citep{Brown2014sousa}.

\subsection{SN 2011by}
\begin{figure}[h]
\plotone{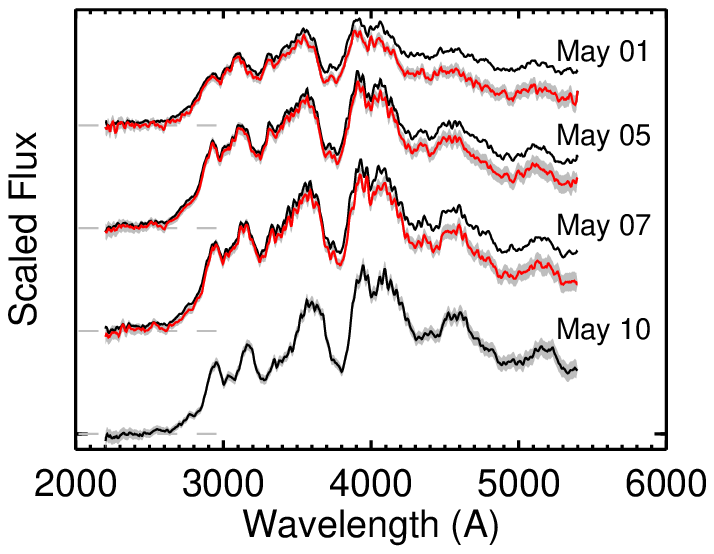} 
\caption{Spectra of SN 2011by extracted using the regular UVOTPY extraction and our empirical decontamination technique.}
\label{2011byspec}
\end{figure}

Four spectra of SN 2011by extracted using our empirical decontamination technique are presented in Figure \ref{2011byspec}.  The phases of these observations were calculated using the UVOT photometry B-band time of maximum light of $JD\,=\,2455690.9$ from SOUSA \citep{Brown2014sousa}.

\subsection{SN 2011fe} 
\begin{figure}[h]
\plotone{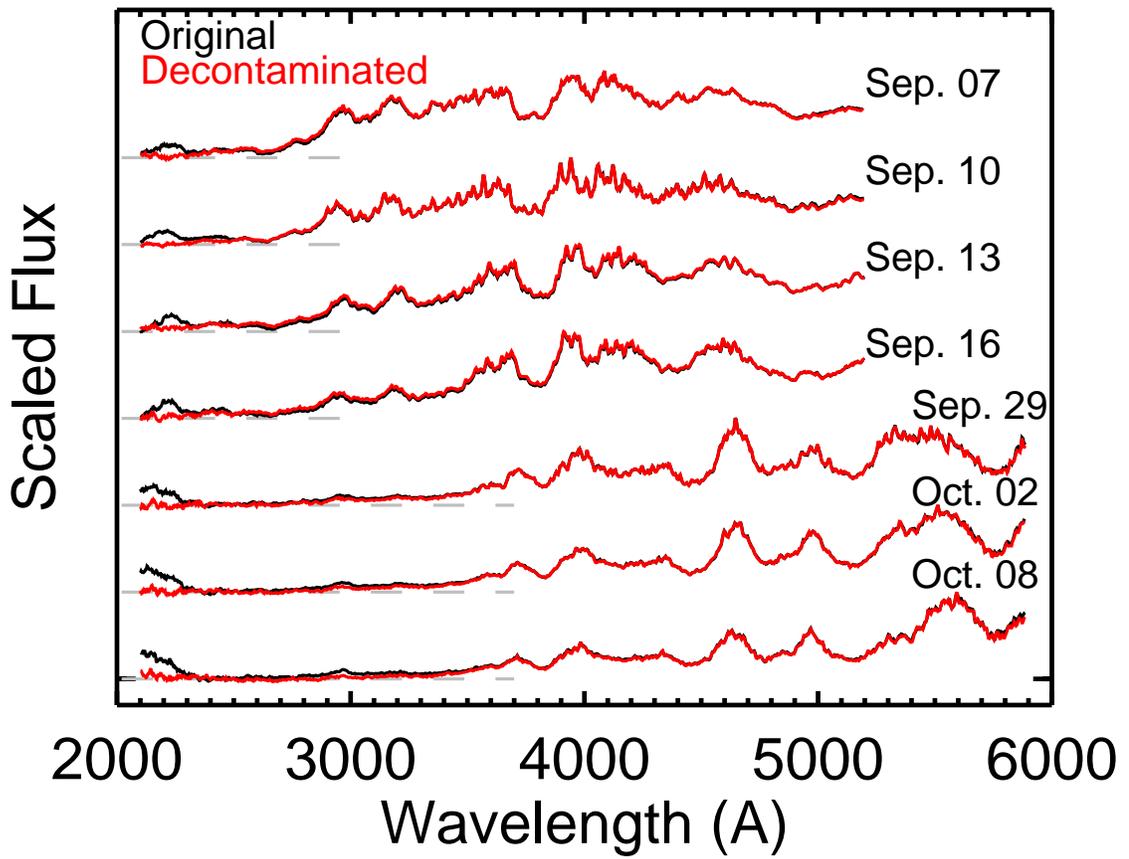} 
\caption{Spectra of SN 2011fe extracted using the regular UVOTPY extraction and our empirical decontamination technique.  Error contours are smaller than the thickness of the plotting lines.}
\label{2011fespecappendix}
\end{figure}

We extracted seven spectra of SN 2011fe using our empirical decontamination technique.  These are presented in Figure \ref{2011fespecappendix}.  The spectrum from 2011 September 10 is the same as in Section \ref{section2011fe}.  We calculated phases relative to B-band maximum light using the peak value of \citet{Pereira2013} occurring at $JD\,=\,2455815.0$. 

\subsection{SN 2012cg}
\begin{figure}[h]
\plotone{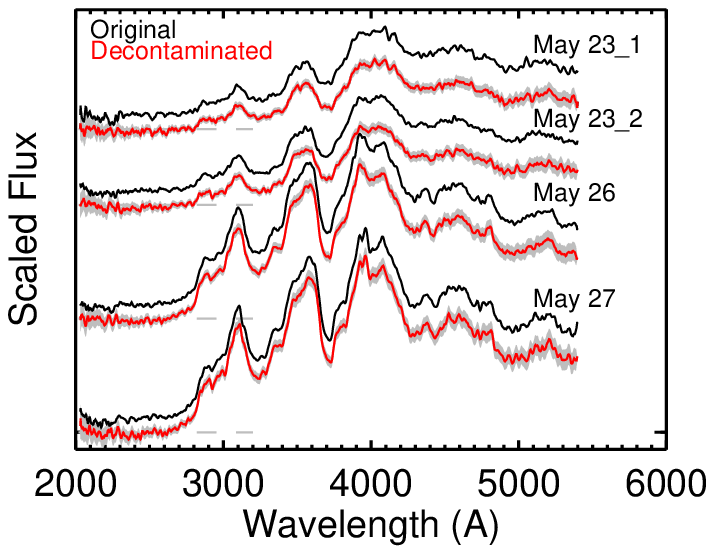} 
\caption{Spectra of SN 2012cg extracted using the regular UVOTPY extraction and our empirical decontamination technique.  The offset between the spectra is due to the first order dispersed host galaxy light being superimposed on the dispersed SN spectrum and not appearing within the background estimation region.}
\label{2012cgspec}
\end{figure}

Four spectra of SN 2012cg extracted using our empirical decontamination technique are presented in Figure \ref{2012cgspec}.  Both obsids contained many exposures, so we broke each into two groups for analysis to give better time resolution.  The phases of these observations were calculated using the UVOT photometry B-band time of maximum light of $JD\,=\,2456082.1$ from SOUSA \citep{Brown2014sousa}.

\subsection{SN 2012dn} 
\begin{figure}[h]
\plotone{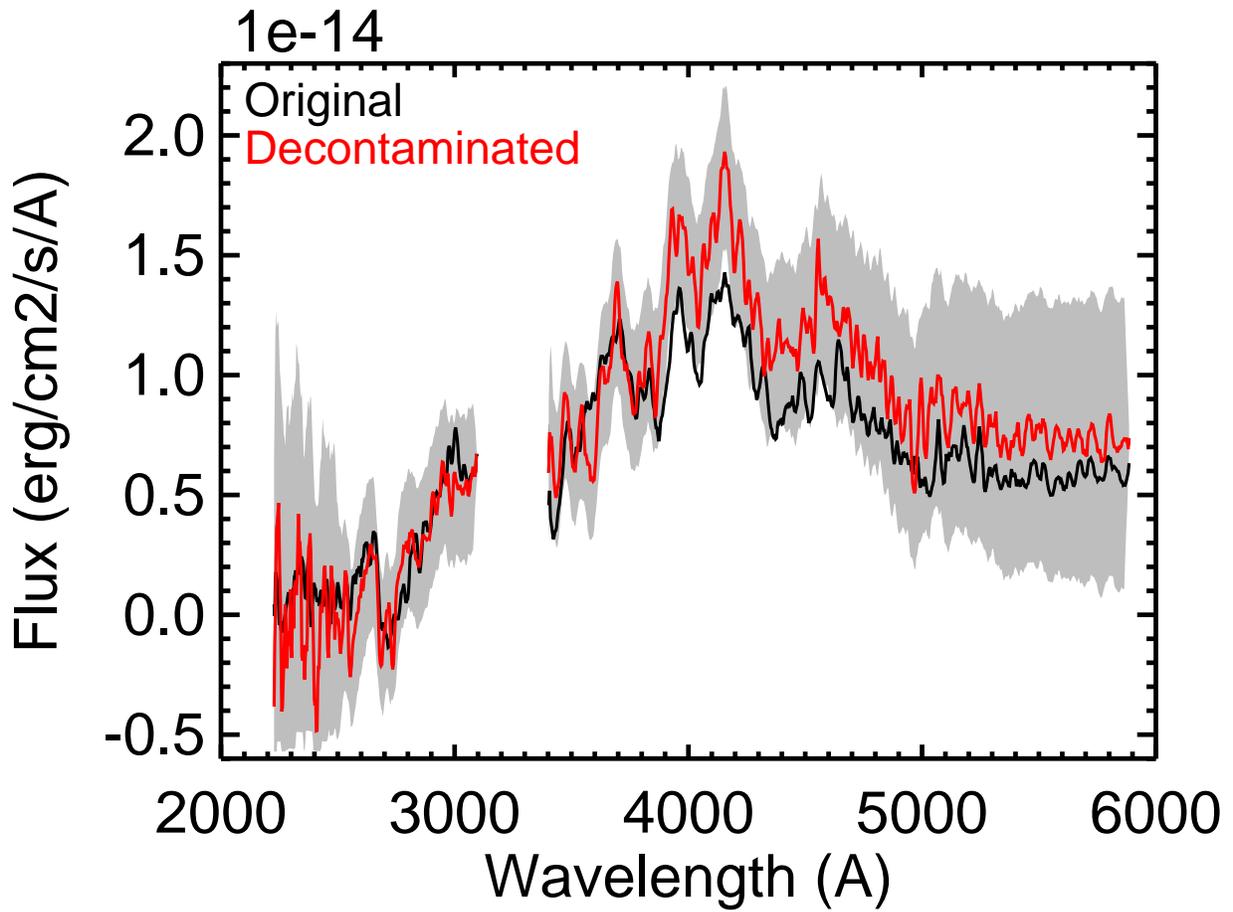} 
\caption{A spectrum of SN 2012dn from 2012 July 22 extracted using the regular UVOTPY extraction and our empirical decontamination technique.  This spectrum was originally presented by \citet{Brown2014}.  A portion of the spectrum was trimmed due to severe contamination from a very bright field source.}
\label{2012dnspec}
\end{figure}

 A UVOT spectrum of SN 2012dn was presented in \citet{Brown2014}.  We re-extracted this spectrum using our empirical decontamination technique and present it in Figure \ref{2012dnspec}.  The spectral features blueward of $2700$\,\AA~ documented by \citet{Brown2014} are still present following decontamination.   The phase of this observation was calculated using the UVOT photometry B-band time of maximum light of $JD\,=\,2456133.3$ \citep{Brown2014}.

\bibliographystyle{apj}

\begin{table}[h]
	\begin{center}
	\tablecomments{B-band maximum light occurred at MJD$=56822.5\pm 0.3$. (Smitka et al., in prep.)}	
	\caption{iPTF14bdn Observation Details \label{14bdntable}}	
	\label{14bdntable}
	\begin{tabular}{cccc}
	\toprule\toprule
        SN Obsid & Template Obsid & Roll Angle & SN Phase \\
                       &                          & (Deg)         & (Days) \\
        \midrule
	00033311012 & 00092182002 & 316 & -6 \\
	00033311015 & 00092182002 & 316 & -3 \\
	00033311021 & 00092182004 & 305 & +5 \\
	00033311025 & 00092182004 & 305 & +9 \\
	\bottomrule
	\end{tabular}
	\end{center}
\end{table}

\begin{table}[h]
	\begin{center}
	\tablecomments{Ellipsis symbols (\ldots) denote spectra for which we did not apply the empirical decontamination technique in the extraction process.  These spectra did not show any signs of contamination and are included for completeness. \\ $^\star$ Relative to B-band maximum light.  \\ $^{\dagger}$ This epoch's observations were comprised of three roll angles, additional template obsids are 00092172006 and 00092172008. }	
	\caption{Observation Details for SNe Ia \label{appendixtable}}	
	\label{appendixtable}
	\begin{tabular}{cccccrr}
	\toprule\toprule
        SN & Obsid & Template Obsid & Date &  JD & Phase$^\star$ & Exposure Time\\
             &           &                          &         & $(2450000+)$ & (days) & (sec) \\
        \midrule
	2005df & 00030252004 & 00092169002 & 2005 Aug 11       & 3599.2 & -5.2  & 1640\\
	2005df & 00030252010 & 00092169002 & 2005 Aug 14       & 3596.6 & -2.2  & 986\\
	2005df & 00030252014 & 00092169002 & 2005 Aug 17       & 3600.4 & 1.6    & 2018\\
	2005df & 00030252019 & \ldots              & 2005 Aug 21      & 3604.1  & 5.2    & 422\\
	2009dc & 00031405005 & 00092171002 & 2009 May 01       & 4952.6 & 5.2    & 7682\\
	2009ig & 00031473007 & 00092172004$^{\dagger}$ & 2009 Aug 25  & 5069.1 & -11.2 & 4548 \\
	2009ig & 00031473010 & 00092172014 & 2009 Aug 27      & 5071.1  & -9.3   & 14019 \\
	2009ig & 00031473014 & 00092172014 & 2009 Sep 01       & 5076.1  & -4.2   & 5853 \\
	2009ig & 00031473017 & 00092172014 & 2009 Sep 03       & 5078.2  & -2.2   & 13524 \\
	2009ig & 00031473021 & 00092172014 & 2009 Sep 07       & 5082.0  & 1.6     & 17337 \\
	2009ig & 00031473026 & 00092172012 & 2009 Sep 14      & 5089.0   & 8.6     & 18010 \\
	2010ev & 00031751001 & 00092173002 & 2010\, Jul \,\,05 & 5383.3  & -2.2  & 17670\\
	2011by & 00031977006 & 00092177002 & 2011 May 01       & 5683.1 & -7.9   & 9451\\
	2011by & 00031977014 & 00092177002 & 2011 May 05       & 5686.0 & -3.9   & 7780\\
	2011by & 00031977019 & 00092177002 & 2011 May 07       & 5686.8 & -2.2   & 9238\\
	2011by & 00031977024 & \ldots               & 2011 May 10      & 5689.0 & 1.1    & 9553\\
	2011fe  & 00032094001 & 00032094018 & 2011 Sep 07       & 5811.7 & -3.2   & 3525\\
	2011fe  & 00032094004 & 00032094018 & 2011 Sep 10       & 5815.0 & 0.1     & 4165\\
	2011fe  & 00032094010 & 00032094018 & 2011 Sep 13       & 5818.0 & 3.0     & 3019\\
	2011fe  & 00032094012 & 00032094018 & 2011 Sep 16       & 5820.6 & 5.7     & 3071\\
	2011fe  & 00032101004 & 00032101012 & 2011 Sep 29       & 5833.8 & 18.7   & 2540\\
	2011fe  & 00032101006 & 00032101012 & 2011 Oct 02       & 5837.4 & 22.3   & 4141\\
	2011fe  & 00032101009 & 00032101012 & 2011 Oct 08       & 5842.9 & 27.8   & 3729\\
	2012cg & 00032464002 & 00092179004 & 2012 May 23       & 6070.7 & -11.4 & 8875\\
	2012cg & 00032464002 & 00092179004 & 2012 May 23       & 6071.1 & -10.9 & 8801\\
	2012cg & 00032464008 & 00092179004 & 2012 May 26       & 6074.0 & -8.1   & 8231\\
	2012cg & 00032464008 & 00092179004 & 2012 May 27       & 6074.9 & -7.2   & 7346\\
	2012dn & 00032516006 & 00092180002 & 2012\, Jul\,\, 22 & 6131.5 & -2.2   & 3574\\

	\bottomrule
	\end{tabular}
	\end{center}
\end{table}

\end{document}